\documentclass[
 reprint,
 amsmath, amssymb,
 pre,
 aps,
 superscriptaddress
]{revtex4-2}

\usepackage{graphicx}
\usepackage{subfigure}
\usepackage{hyperref}
\usepackage{bm}
\usepackage{float}
\usepackage{xcolor}
\usepackage{amsmath}
\usepackage{braket} 
\usepackage{orcidlink}

\DeclareMathOperator{\Tr}{Tr}

\newcommand\Expval[1]{\left\langle{#1}\right\rangle}

\newcommand\abs[1]{\vert{#1}\vert}

\definecolor{midnightblue}{RGB}{0, 0, 128}
\definecolor{mediumblue}{RGB}{0, 0, 205}
\definecolor{darkseagreen}{RGB}{66, 169, 87}
\definecolor{darkred}{RGB}{205,92,92}
\definecolor{deepskyblue}{HTML}{00BFFF}
\definecolor{goldenrod}{RGB}{218,165,32}  

\hypersetup{colorlinks=true, citecolor=blue, urlcolor=blue, linkcolor=blue}

\begin{document}

\preprint{APS/123-QED}

\title{Entanglement-enabled Criticality in  One-dimensional Quantum Contact Process}

\author{Ya-Xin Xiang}
\altaffiliation{These authors contributed equally to this work.}
\affiliation{National Laboratory of Solid State Microstructures and School of Physics, Collaborative Innovation Center of Advanced Microstructures, Nanjing University, Nanjing 210093, China}

\author{Tianyi Yan}
\altaffiliation{These authors contributed equally to this work.}
\affiliation{School of Physics and Astronomy, and Centre for the Mathematics and Theoretical Physics of Quantum nonequilibrium Systems, University of Nottingham, Nottingham, G7 2RD, UK}

\author{Weibin Li}
\email{weibin.li@nottingham.ac.uk}
\affiliation{School of Physics and Astronomy, and Centre for the Mathematics and Theoretical Physics of Quantum nonequilibrium Systems, University of Nottingham, Nottingham, G7 2RD, UK}

\author{Yu-Qiang Ma}
\email{myqiang@nju.edu.cn}
\affiliation{National Laboratory of Solid State Microstructures and School of Physics, Collaborative Innovation Center of Advanced Microstructures, Nanjing University, Nanjing 210093, China}
\affiliation{Hefei National Laboratory, Hefei 230088, China}
     
\begin{abstract}
The contact process is a paradigmatic example of nonequilibrium dynamics, with broad applications ranging from chemistry to sociology. Its quantum counterpart—the quantum contact process (QCP)—extends the classical model to include coherent processes. Despite sustained interest, the nature of the transition in the one-dimensional (1D) QCP remains debatable. Here, combining Liouvillian spectral analysis, the tensor jump method, exact quantum jump Monte Carlo, and truncated Wigner simulations, we show that 1D QCP undergoes a continuous absorbing-state phase transition, with critical exponents distinct from the classical case. We further find Liouvillian gap closes well below the critical point, highlighting that spectral gap analysis alone cannot distinguish a phase transition from metastability in the QCP. Crucially, the 1D QCP is weakly entangled—yet even this weak entanglement is indispensable for capturing the correct critical behavior, whereas semiclassical methods artificially stabilize the active state and predict a spurious first-order transition.
Our work establishes the quantum origin of the phase transition in the 1D QCP and underscores the essential role of entanglement in dissipative quantum many-body systems.
\end{abstract}
\maketitle

\textit{Introduction.—}Nonequilibrium many-body dynamics in driven-dissipative systems is a topic of fundamental interest. Such systems generally cannot be studied via traditional methods of equilibrium statistical mechanics, although core concepts—criticality, scaling, and universality—still apply. A paradigmatic example is the contact process on a lattice. In its classical form, each site is either active or inactive, and the dynamics is driven by incoherent branching and decay. The classical contact process has broad applications in chemistry, biology, and sociology~\cite{Albano1994Forest,tauber2024Stochastic,lindner2004effects,Kerr2002Local,Szabo2016Evolutionary,Satulovsky1994Stochastic}, and exhibits an absorbing-state phase transition (APT) between an absorbing state and an active one, with critical behavior belonging to the directed percolation (DP) universality class~\cite{janssen1981nonequilibrium,grassberger1981phase,dickman1991time,Antal2001Phase,henkel2009noneq}.
Its quantum counterpart—the quantum contact process (QCP)—promotes classical branching and coagulation to coherent processes, in which an inactive site is activated by its active neighbors via either coherent (quantum) or dissipative (classical) channels~\cite{marcuzzi2016absorbing,Buchhold2017Nonequilibriumeffectivefield,marcuzzi2015non,Bohm2026Quantum}. 

While the classical regime of the QCP features a well-established continuous APT~\cite{epidemic2017perez,gutierrez2017exp,helmrich2020signatures,brady2025anomalous}, the order of the phase transition in the quantum regime remains fiercely debated. Mean-field analysis reveals a first-order APT~\cite{shang2026steady,marcuzzi2016absorbing}, whereas field-theoretical approaches argue that strong fluctuations soften it into a continuous transition in one dimension (1D)~\cite{roscher2018phe,xiang2023self}. Numerical studies are also divided. Small-system quantum-jump Monte Carlo simulations (QJMC) reveal a bimodal order-parameter distribution reminiscent of a first-order transition~\cite{marcuzzi2016absorbing}. In contrast, infinite time-evolving block decimation and its trajectory-based variant point to a continuous transition that belongs to a different universality class than DP~\cite{carollo2019critical,Gillman2019Numerical,Jebai2026Tuning}. A recent study combining Liouvillian spectral analysis with cluster mean-field approaches has further revived the first-order scenario by claiming a saddle-node bifurcation of the order parameter~\cite{shang2026steady}. 

This controversy raises a fundamental question: how quantum is the QCP? Recent work has recognized the potential role of quantum correlations—coherence and entanglement—in APTs~\cite{Odea2024entanglement,gillman2020noneq,gillman2021quantum,carollo2022dark}. Their precise impact on the QCP remains to be determined.
A further layer of complexity arises from the spectral properties of the QCP. In finite systems, the decrease of the Liouvillian gap with system size is often interpreted as a precursor of a phase transition in the thermodynamic limit~\cite{minganti2018spectral,foss2017emergent,kessler2012dissipative,Xiang2026Swicth,li2024collective}. In the presence of an absorbing state, however, gap closing is necessary but not sufficient: because the steady state is uniquely determined by the absorbing state, the gap may close even when no true transition occurs——as in a birth-death process, where the gap closing merely reflects the increasing rarity of extinction event with system size~\cite{kamenev2008extinction,park2017extinction,lohmar2011switching}.

In this work, we settle the controversy through a comprehensive numerical study using Liouvillian spectral analysis, the tensor jump method (TJM)~\cite{Sander2025}, exact QJMC~\cite{plenio1998the,gardiner1992wave,dalibard1992wave}, and the open-system discrete truncated Wigner approximation (OSDTWA)~\cite{singh2022driven}. This multi-method approach allows us to perform the first systematic finite-size analysis of the 1D QCP and leads to three central results. First, we show that the closing of the Liouvillian gap with system size does not itself signal an APT, as the threshold for power-law gap closing lies well below the true critical point. Second, we extract the critical exponents and find that they differ from the DP. 
Finally, and most importantly, we uncover the decisive role of weak quantum entanglement in the critical behavior: even weak and short-ranged entanglement suffices to capture the true critical dynamics, whereas the first-order transition predicted by semiclassical methods is an artifact of truncation. Away from criticality, the order parameter dynamics decouple from entanglement, as demonstrated by the agreement between semiclassical and full quantum treatments. Together, these results resolve the long-standing debate, establish the transition as genuine quantum critical phenomenon, and highlight entanglement as the essential ingredient for quantum criticality for dissipative systems.

\begin{figure}[t]
    \centering
    \includegraphics[width = 8.6cm, keepaspectratio]{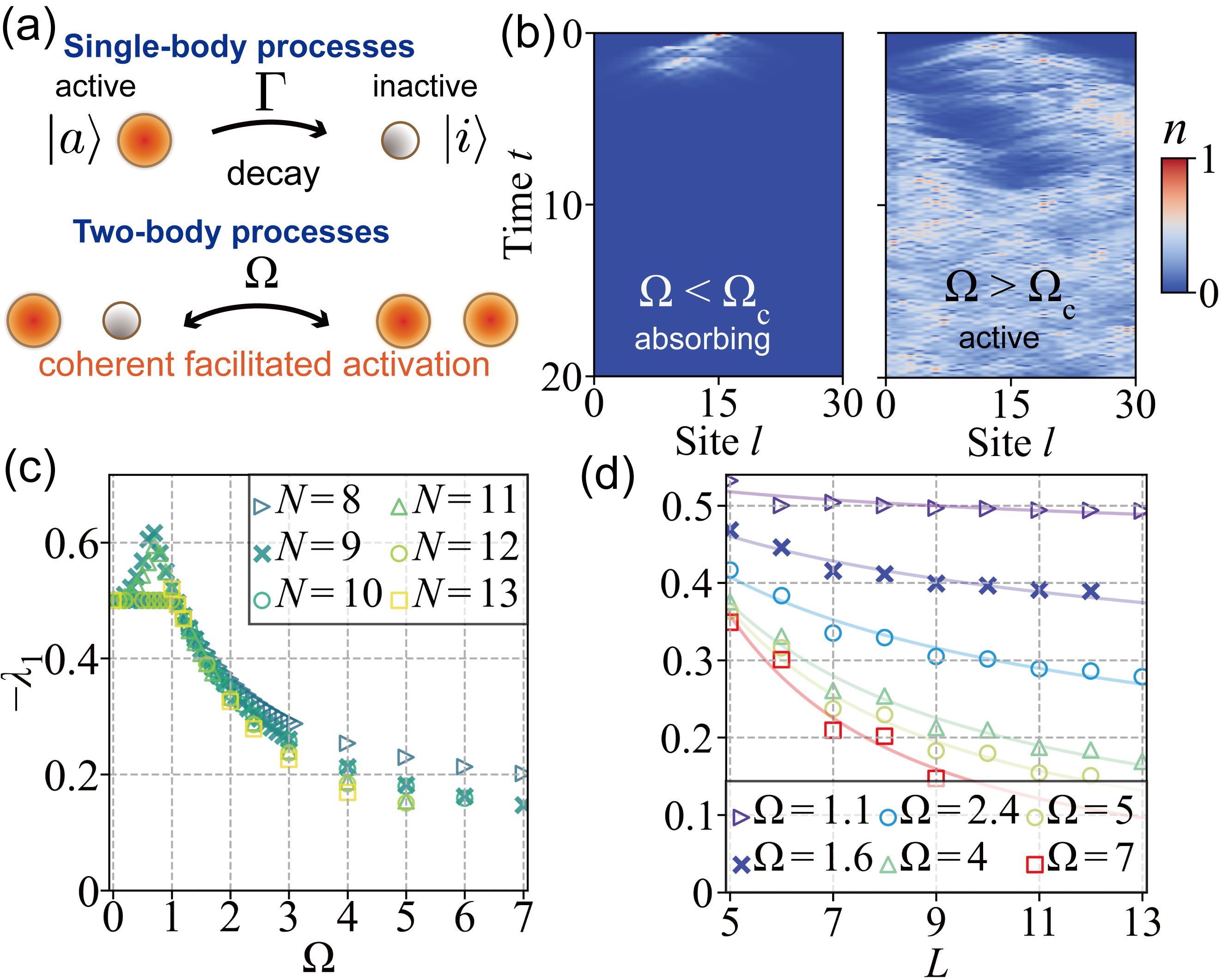}
    \caption{(a) Schematic of the quantum contact process.
    An inactive site (state $\ket{i}$, gray sphere) can be activated via coherent facilitated activation by a neighboring active state $\ket{a}$ (red sphere) at rate $\Omega$. Active sites spontaneously decay back to $\ket{i}$ at rate $\Gamma$.
    (b) Spatiotemporal profile of the active density $n$ for different facilitation rates $\Omega$. Starting at a fully active state, small $\Omega<\Omega_c$ leads to a rapid decay into the absorbing state, while large $\Omega>\Omega_c$ sustains a long-lived active state. The critical $\Omega_c=6.5$ marks the boundary between the two dynamical regimes (see details in the main text).
    (c) Spectral gap versus $\Omega$ for different system sizes $N$. The gap decreases with both increasing $\Omega$ and $N$ beyond a certain threshold $\Omega\approx1$, which is far below the true critical point $\Omega_c=6.5$.
    (d) Finite-size scaling of the gap for various $\Omega$. Solid lines represent power-law fits $-\lambda_1\propto L^{-b}$.}    
    \label{fig:scheme}
\end{figure}

\textit{Model.—}We consider a generic QCP on a 1D lattice of $N$ sites with unit spacing and periodic boundary conditions. As depicted in Fig.~\ref{fig:scheme}(a), each site is either active ($\ket{a}$) or inactive ($\ket{i}$)~\cite{marcuzzi2016absorbing}. The active site can spontaneously become inactive at rate $\Gamma$, and the inactive one can be coherently activated by neighboring active ones at rate $\Omega$.
Under the Markovian noise, the dynamics is governed by a Lindblad master equation for the density operator $\hat\rho$ ($\hbar = 1$),
\begin{equation}
\hat{\mathcal{L}}[\hat\rho]\equiv\partial_t \hat\rho = -i[\hat H,\hat\rho] + \sum_{l} \mathcal{D}_{l}[\hat\rho].
\end{equation}
The coherent activation is described by the Hamiltonian,
\begin{equation}
	\hat H =\Omega\sum_l\hat C_l \hat\sigma_l^x,~\text{with}~
    \hat C_l=\sum_{k\in\partial_l}\hat n_k.
\end{equation}
Here, $\Omega$ is the facilitated activation rate, and $k$ and $l$ label individual sites and the summation $\sum_{k\in\partial l}$ runs over the effective nearest neighbors of the $l$-th site. 
We define the local transition operators ${\hat\sigma}_l^{\alpha\beta}\equiv\ket{\alpha_l}\bra{\beta_l}$ ($\alpha, \beta= a, i$), the activation number operator $\hat n_l=\hat\sigma^{aa}_l$,  and the Pauli operator ${\hat\sigma}_l^x=\hat\sigma^{-}_l+\hat\sigma^{+}_l$ and $\hat\sigma^{y}_l=i({\hat\sigma}_l^{-}-{\hat\sigma}_l^{+})$. The latter serves to flip the quantum state via the ladder operators $\hat\sigma^{+}_l \equiv \hat\sigma_l^{ai}$ and $\hat\sigma^{-}_l \equiv \hat\sigma_l^{ia}$. 

Local dissipative process is given by the superoperator, 
\begin{equation}
    \mathcal{D}_{l}[\hat\rho] ={\hat L}_{l}\hat\rho {\hat L}_{l}^\dagger -\frac{1}{2} \left\{{\hat L}_{l}^{\dagger}{\hat L}_{l},\hat\rho \right\},
\end{equation}
where $\hat L_l = \sqrt{\Gamma}\hat\sigma_l^-$ is the jump operator for spontaneous decay of active site $l$. Throughout this work, we scale time in unit of $\Gamma^{-1}=1$. We use the active density $n\equiv\sum_l\Tr{[\hat\rho \hat n_l]}/N$ as the order parameter, where an active (absorbing) state is defined as a state with finite (zero) active density. As the facilitation rate increases, the system undergoes an APT: for strong facilitation, the system settles into an active state, while for weak facilitation, it decays rapidly into the absorbing state, as illustrated in Fig.~\ref{fig:scheme}(b).

\textit{Liouvillian spectrum.—}We first investigate spectral signatures of QCP. We obtain its eigenvalues and eigenmatrices via $\hat{\mathcal{L}}[\hat{\rho}_l]=\lambda_l\hat{\rho}_l$ and sort them according to their real parts, $0=\lambda_0\geq\Re[\lambda_1]\geq...$. It follows from $\lambda_0=0$ that the stationary density matrix $\hat{\rho}_\text{ss}=\hat{\rho}_0/\Tr{[\hat{\rho}_0]}$, and the Liouvillian spectral gap is defined as $-\Re[\lambda_1]=-\lambda_1$, since $\lambda_1$ is real for the parameter considered. It is straightforward to verify that the all-inactive state—the absorbing state—is an eigenstate of the Lindbladian with zero eigenvalue.
Since the Lindbladian typically has a unique steady state, the absorbing state is the stationary state, $\hat\rho_\text{ss}=\prod_l\hat\sigma_l^{ii}$, independent of parameters, and the system cannot escape from the absorbing state once it is reached~\cite{Haye2006nonequilibrium}.

Fig.~\ref{fig:scheme}(c) shows the spectral gap as a function of $\Omega$ for various system sizes $N$. The gap starts to decrease with increasing $N$ and $\Omega$ beyond $\Omega \approx 1$. In Fig.~\ref{fig:scheme}(d), the gap shows a power-law dependence on the system size, $-\lambda_1\propto L^{-b}$~\cite{shang2026steady,carollo2019critical}, indicating a gap closing in the thermodynamic limit. Yet, previous works have drawn conflicting conclusions from this algebraic closing of spectral gap——some inferring a first-order APT~\cite{shang2026steady}, others a continuous one~\cite{carollo2019critical}. This ambiguity underscores that gap closing alone cannot distinguish the order of APT. 
We address this ambiguity by examining the long-time quantum dynamics, which goes beyond the spectral analysis.

\begin{figure}[t]
    \centering
    \includegraphics[width = 8.6cm, keepaspectratio]{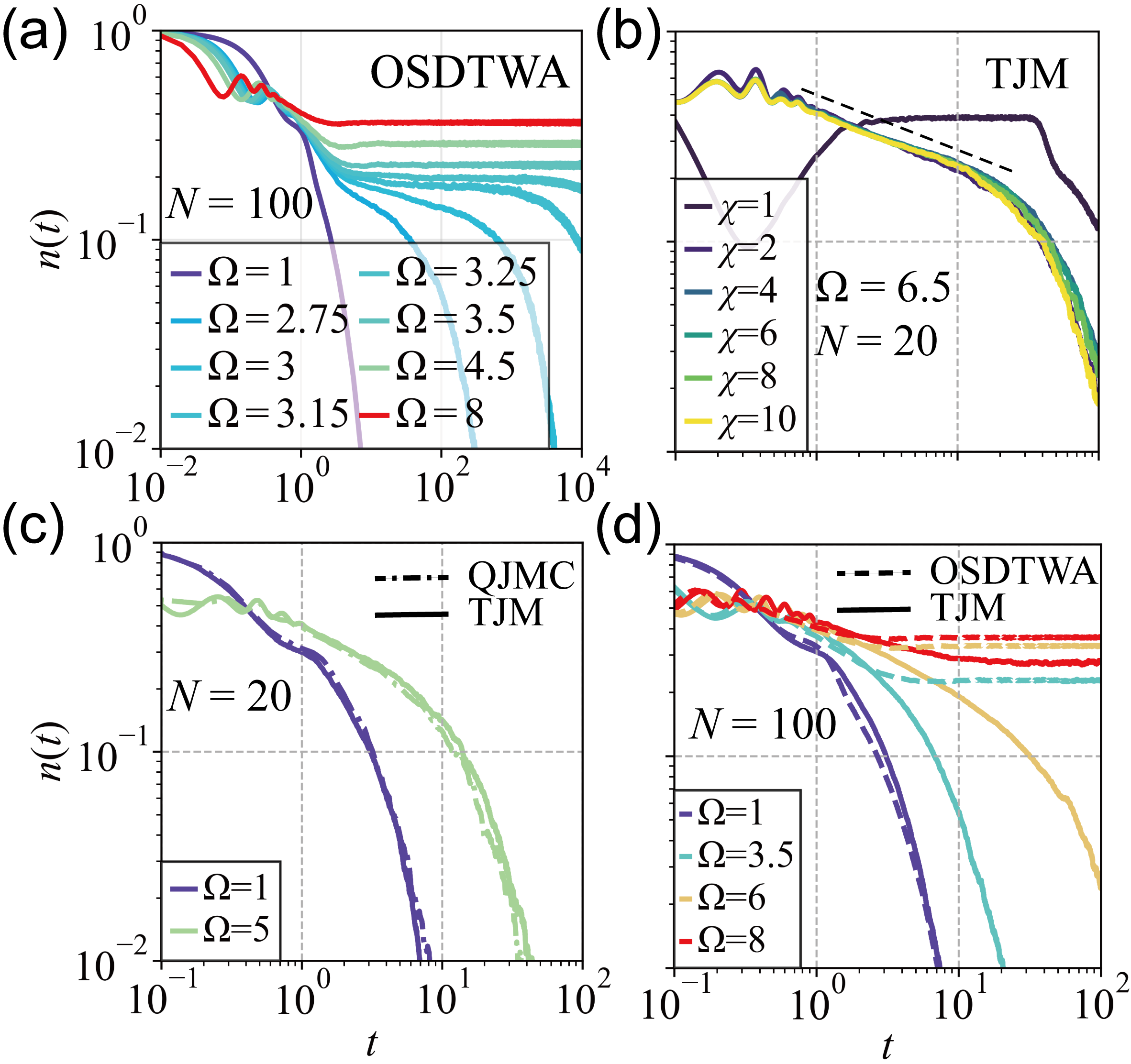}
    \caption{Decay dynamics of the active density $n(t)$ obtained from different methods. 
    (a) Semiclassical open-system discrete truncated Wigner approximation (OSDTWA) results for $N=100$ at various facilitation rates $\Omega$. For $\Omega\leq3.15$, $n(t)$ decays to the absorbing state. For $\Omega\geq3.15$, a long-lived active plateau emerges. 
    (b) Convergence of the tensor jump method (TJM) with various bond dimensions $\chi$ at $\Omega=6.5$ and $N=20$. The black dashed line marks a power-law decay.
    (c) Benchmark of TJM ($\chi=4$) against exact quantum-jump Monte Carlo (QJMC) simulations for $\Omega=1,5$ and $N=20$.
    (d) Comparison of TJM (solid lines) and OSDTWA (dashed lines) for $N=100$ at four representative values of $\Omega$. 
    }    
    \label{fig:compare}
\end{figure}
\textit{Quantum effects.—}
We investigate 1D QCP using well-established methods, including OSDTWA, TJM, and exact QJMC. 
While OSDTWA retains quantum coherence and fluctuations up to second order but neglects entanglement, TJM systematically captures entanglement through its matrix-product-state representation, with bond dimension $\chi$ controlling the amount of entanglement ($\chi=1$ corresponds to zero entanglement). QJMC serves as an exact reference on small systems, free from both truncation and semiclassical approximations. This combination allows us to isolate the role of entanglement and correlation in the transition.

As shown in Fig.~\ref{fig:compare}(a), the semiclassical dynamics from OSDTWA starting from a fully active state exhibits a rapid decay to the absorbing state below $\Omega=3.15$ and a long-lived active plateau above it, with no power-law decay in between——a hallmark of a first-order transition within the semiclassical approximation. This raises the question of whether such a first-order-like behavior persists when quantum entanglement is properly accounted for.

To investigate this thoroughly, we turn to TJM. We first demonstrate the convergence of TJM with bond dimension $\chi$ at $\Omega=6.5$ and $N=20$. The results, shown in Fig.~\ref{fig:compare}(b), reveal that neglecting entanglement ($\chi=1$) produces a plateau reminiscent of OSDTWA. In contrast, simulation results for $\chi\geq 2$ collapse, apart from a small early-time deviation at $\chi=2$, indicating that $\chi=4$ already suffices to capture the relevant quantum correlations. Notably, these curves exhibit an intermediate-time power-law decay (black dashed line), suggestive of critical dynamics enabled by weak quantum entanglement. To further validate TJM, we benchmark TJM with $\chi=4$ against exact QJMC simulations for $N=20$ at $\Omega=1$ and $5$, shown in Fig~\ref{fig:compare}(c); the two methods show excellent agreement. 

\begin{figure}[b]
    \centering
    \includegraphics[width = 8.6cm, keepaspectratio]{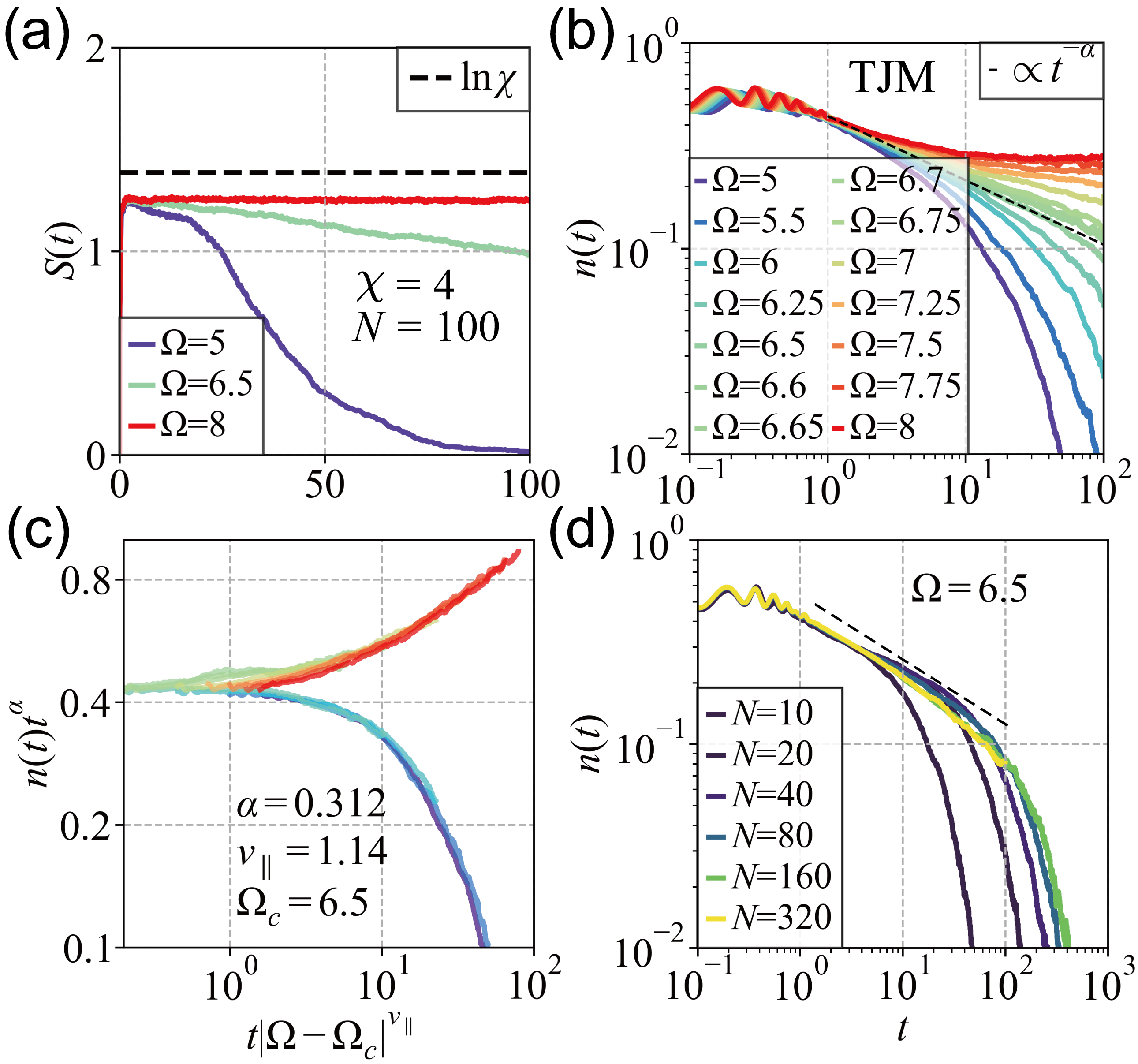}
    \caption{Decay dynamics starting with a fully active state obtained from TJM. 
    (a) Entanglement entropy $S(t)$ and 
    (b) active density $n(t)$ for $N=100$ at various $\Omega$. A power-law decay $n(t)\propto t^{-\alpha}$ (black dashed line) emerges at $\Omega=6.5$, yielding the critical point $\Omega_c=6.5$ and the exponent $\alpha=0.312$.
    (c) Determination of the exponent $\nu_\parallel$ from data collapse of $n(t)t^{\alpha}$ versus $t\abs{\Omega-\Omega_c}^{\nu_\parallel}$. With $\nu_\parallel=1.14$, all curves collapse onto two different branches depending on the sign of $\Omega-\Omega_c$.
    (d) active density $n(t)$ at $\Omega=6.5$ at various system size $N$. As $N$ increases, the curves converge to the universal power law $n(t)\propto t^{-\alpha}$ (black dashed line).
    }    
    \label{fig:decay_tjm}
\end{figure}

Having established the reliability of TJM, we now compare it with OSDTWA for a range of $\Omega$. 
As shown in Fig.~\ref{fig:compare}(d), away from the transition regime ($\Omega=1,8$, the two methods agree qualitatively. However, in the intermediate regime ($\Omega=3.5, 6$), they exhibit a qualitatively different behavior: OSDTWA predicts a plateau, while TJM produces a continuous decay. 
To understand the origin of this discrepancy, we calculate the entanglement entropy $S(t)$ for $\Omega=5, 6.5$ and $8$, displayed in Fig.~\ref{fig:decay_tjm}. In all cases, $S(t)$ initially grows from zero. At $\Omega=5$, it decays rapidly as the system enters the absorbing state. At $\Omega=6.5$, the decay is significantly slower. At $\Omega=8$, it saturates to a plateau. This means that the dynamics of $n(t)$ is less sensitive to entanglement in the strongly driven regime than near criticality. Moreover, the entropy is bounded by $\ln\chi$ (black dashed line), confirming that $\chi=4$ already captures the relevant correlations.

The contrasting sensitivity of $n(t)$ to entanglement admits a simple interpretation. In the deep active regime (large $\Omega$), the system resides in a half-active manifold, where different quantum states—entangled or not—yield the same active density, which is why OSDTWA agrees with TJM. Near criticality, however, the system must explore the Hilbert space between the active and absorbing states. Here, entanglement provides coherent channels through which the active state can continuously leak toward the absorbing state. A product-state ansatz removes these channels entirely, the system can then only escape the active state via a classical nucleation event, yielding a first-order transition. These results demonstrate that even though the system is only weakly entangled, such entanglement is essential for capturing the correct critical dynamics.

\begin{figure}[t]
    \centering
    \includegraphics[width = 8.6cm, keepaspectratio]{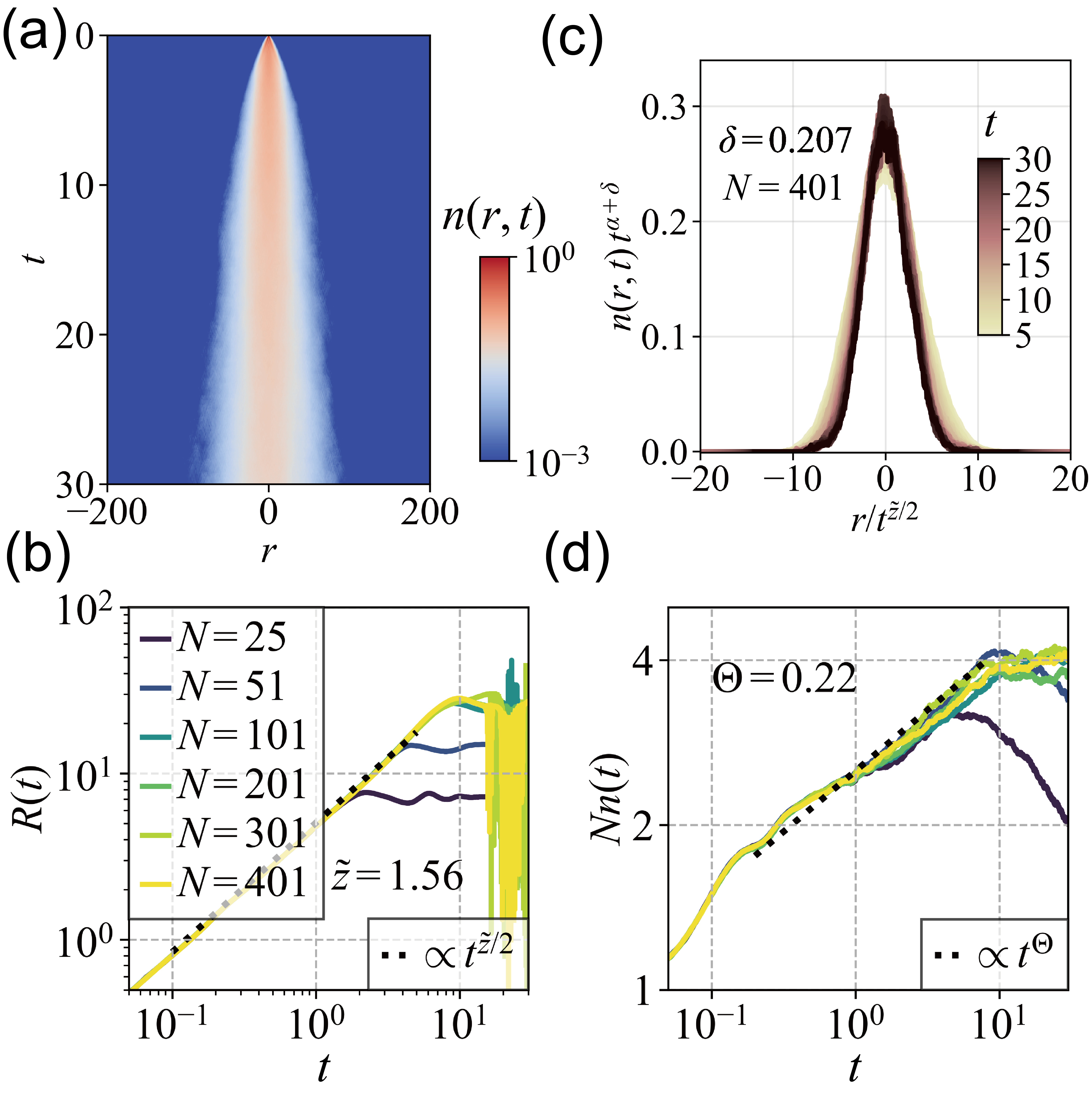}
    \caption{Spreading dynamics starting from a seed initial state at the center of the chain, obtained from TJM at the critical point $\Omega=6.5$. 
    (a) Ensemble average of the active density $n(r,t)$ for $N=401$, where $r\in[-N/2,N/2]$ is the distance from the seed. 
    (b) Mean-squared displacement $R(t)$ of the active cluster for different system sizes $N$. All curves exhibit a universal power-law growth $R(t)\propto t^{\tilde z/2}$ (black dashed line), yielding the exponent $\tilde z=1.56$.
    (c) Collapse of the spatial profiles $n(r,t)$ at different times $t$ (color-coded) for $N=401$. The profiles collapse onto a single curve when plotted as $n(r,t)t^{\alpha+\delta}$ versus $r/t^{\tilde z /2}$, yielding the exponent $\delta=0.203$. 
    (d) Total active density $Nn(t)=\sum_r n(r,t)$ for various $N$ exhibit a power-law growth $Nn(t)\propto t^\Theta$ (dashed black line) with exponent $\Theta=0.22$. }    
    \label{fig:seed_tjm}
\end{figure}

\textit{Dynamical scaling.—}With the critical behavior established, we now perform a systematic dynamical scaling analysis~\cite{henkel2009noneq,Grassberger1979Reggeon}, using TJM with $\chi=4$.
Fig.~\ref{fig:decay_tjm}(b) shows decay dynamics of the active density $n(t)$. For $N=100$, as $\Omega$ increases from $5$ to $8$, $n(t)$ transitions from a rapid decay to the absorbing state at small $\Omega$ to an active plateau at large $\Omega$. A power-law decay $n(t)\propto t^{-\alpha}$ emerges at $\Omega=6.5$, indicating the critical point $\Omega_c=6.5$ and yielding the exponent $\alpha=0.312$. Under the dynamical scaling hypothesis for homogeneous active initial state~\cite{henkel2009noneq}, the exponent $\nu_\parallel$ is determined by collapsing $n(t)t^{\alpha}$ against $t\abs{\Omega-\Omega_c}^{\nu_\parallel}$. With $\alpha=0.312$ and $\nu_\parallel=1.14$, all curves collapse onto two distinct branches depending on the sign of $\Omega-\Omega_c$, distinguishing the supercritical (upper) and subcritical (lower) regimes. 
We emphasize that the system size must be sufficiently large for this critical dynamics to emerge. As shown in Fig.~\ref{fig:decay_tjm}(d), as $N$ increases, the curves converge to the universal power law $n(t)\propto t^{-\alpha}$ (black dashed line), while smaller systems exhibit strong critical fluctuations that deviate from the scaling behavior. 

\begin{table}[b]
\centering
\caption{Critical exponents of the QCP from this work, compared with previous estimates (Refs.~\cite{carollo2019critical,Gillman2019Numerical,Jebai2026Tuning}) and directed percolation (DP) in one dimension (1D) and two dimensions (2D)~\cite{Hinrichsen2000NonEquilibrium}. Here $\tilde z = 2/z$.}
\begin{tabular}{c|c|c|c|c}
\hline\hline
 & QCP (this work) &QCP (Refs.~\cite{carollo2019critical,Gillman2019Numerical,Jebai2026Tuning}) & 1D DP & 2D DP  \\
\hline
$\beta$     & 0.36 & 0.39 & 0.28 &0.58\\
$\nu_\perp$ & 0.89 & 0.5 & 1.1 &0.73 \\
$\nu_\parallel$ & 1.14 & 1 & 1.73 &1.3\\
$\tilde z$      & 1.56 & 1.24 & 1.26 & 1.13 \\
$\alpha$        & 0.312 & 0.36 & 0.16 &0.45 \\
$\delta$        & 0.207 & 0.26 & 0.16 &0.45 \\
$\Theta$        & 0.22 & 0.26 & 0.31 & 0.23\\
\hline\hline
\end{tabular}
\label{tab:exponents}
\end{table}

To further extract critical exponents, we now study the spatial spreading dynamics of the active cluster starting from a seeded initial state, as presented in Fig.~\ref{fig:seed_tjm}. The ensemble-averaged spatiotemporal profile of the active density $n(r,t)=\Tr{[\hat\rho(t)\hat n_{r+N/2}]}$, where $r\in[-N/2,N/2]$ is the distance from the seed, is shown in Fig.~\ref{fig:seed_tjm}(a), revealing diffusive spreading from the seed. We characterize the active cluster width  via the mean-squared displacement, 
\begin{equation}
    R(t)=\Expval{\frac{\sum_rn(r,t)r^2}{\sum_r n(r,t)}}^{1/2}.
\end{equation}
As shown in Fig.~\ref{fig:seed_tjm}(b), $R(t)$ for different system sizes $N$ exhibits a common power-law growth $R(t)\propto t^{\tilde z /2}$ (black dashed line), yielding the exponent $\tilde z= 1.56$. By collapsing the spatial profiles of the active density $n(r,t)$ at different times for $N=401$ onto a single curve via plotting $n(r,t) t^{\alpha+\delta}$ against $r/t^{\tilde z/2}$~\cite{Grassberger1979Reggeon}, we extract the exponent $\delta=0.207$, confirming the dynamical scaling behavior of the spreading profiles. Finally, the total active density $N n(t)=\sum_r n(r,t)$ also exhibit universal scaling, with curves for different $N$ exhibiting the power law $N n(t)\propto t^\Theta$, with $\Theta=0.22$, as shown in Fig.~\ref{fig:seed_tjm}(d). From the standard scaling relation~\cite{henkel2009noneq}, we obtain the exponents $\nu_\perp=\tilde z \nu_\parallel/2= 0.89$ and $\beta=\alpha \nu_\parallel = 0.36$. The extract exponents, summarized in Table~\ref{tab:exponents}, differ from those of DP, indicating that quantum entanglement plays a crucial role in shaping the critical behavior.
Moreover, the rapidity-reversal symmetry is violated ($\delta\neq\alpha$), as it is not a generic feature of QCP. Nevertheless, the hyperscaling relation ($\Theta = \tilde z/2-\delta-\alpha$) is nearly satisfied, as it follows from the more general dynamical scaling hypothesis.

\textit{Conclusions and discussions.—}In this work, we have studied the nature of the APT in the 1D QCP by combining Liouvillian spectral analysis, TJM, QJMC, and OSDTWA. Through systematic finite-size scaling analysis,  we have found that the threshold for power-law closing of the spectral gap ($\Omega\approx1$) is significantly smaller than the true critical point ($\Omega=6.5$). This highlights that in open quantum systems with absorbing states, spectral gap analysis alone is insufficient to determine the nature of a phase transition.
Our results unambiguously establish that the 1D QCP exhibits a continuous APT, in agreement with earlier works~\cite{carollo2019critical,Gillman2019Numerical,Jebai2026Tuning} but now supported by the systematic finite-size analysis of this model.

The most striking aspect of our findings is the decisive role played by weak, 
short-ranged entanglement near criticality. A bond dimension of $\chi=4$—corresponds to a maximum entanglement entropy of $\ln4\approx1.39$—already suffices to recover the correct critical dynamics. The apparent first-order behavior predicted by semiclassical approaches—including OSDTWA and recent cluster mean-field studies—is an artifact of truncation: both methods stabilize the active states by either restricting entanglement to finite clusters or dropping it entirely. The spurious discontinuity disappears when entanglement is treated continuously across the entire system. Away from the critical regime, however, entanglement and the order parameter dynamics decouple, explaining why semiclassical methods remain accurate in both the weakly and strongly driven regimes.

Our results also point to several broader directions. The coupled dynamics of entanglement and the order parameter establishes the genuine quantumness of the 1D QCP, suggesting that such coupling may serve as a hallmark of quantum criticality in dissipative systems. An important open question is whether this scenario extends to higher spatial dimensions, and how additional dissipative channels—such as dephasing—destroy this order parameter-entanglement coupling and eventually restore classical DP behavior. The observed violation of rapidity-reversal symmetry ($\delta\neq\alpha$) further calls for a theoretical understanding, for instance within the Keldysh framework~\cite{Sieberer2016Keldysh,Noel2026PathIntegral,gerbino2024large}. The extracted exponents may reflect a distinct universality class for weakly entangled open quantum systems. Whether such entanglement-enabled critical behavior is generic across other quantum APTs—and whether entanglement itself exhibits universal scaling—remains an open question.

\vspace{0.5cm}

\textit{Acknowledgments.}--We thank Igor Lesanovsky and Juan P. Garrahan for insightful discussions and valuable comments on this work. We are grateful for the computational resources provided by the University of Nottingham's Ada High-Performance Computing service and Nanjing University's High-Performance Computing Center.
We acknowledge financial support from the National Natural Science Foundation of China (Grant Nos. 12347102, 12547183), the EPSRC (Grant No. EP/W015641/1), the Natural Science Foundation of Jiangsu Province (Grant No. BK20233001), the Quantum Science and Technology-National Science and Technology Major Project (Grant No. 2024ZD0300101), and the Fundamental and Interdisciplinary Disciplines Breakthrough Plan of the Ministry of Education of China (Grant No. JYB2025XDXM502). 

\vspace{0.5cm}

\textit{Data Availability}——The data that supports the findings in this work is available on
Zenodo~\cite{data}.


%

\end{document}